# **Optimizing growth performance of *Abelmoschus esculentus* (L.) via synergistic effects of biogenic Cu/Ni/Co oxide nanoparticles in conjunction with rice straw and pressmud based vermicompost**

Shalu Yadav, Ankit Singh, Praveen Kumar Srivastava*, Abhay Kumar Choubey

Department of Sciences and Humanities, Rajiv Gandhi Institute of Petroleum Technology, Jais, Amethi-229304, India

*Corresponding author:

E-mail address: pksrivastava@*rgipt.ac.in* (P. K. Srivastava)

## Graphical Abstract

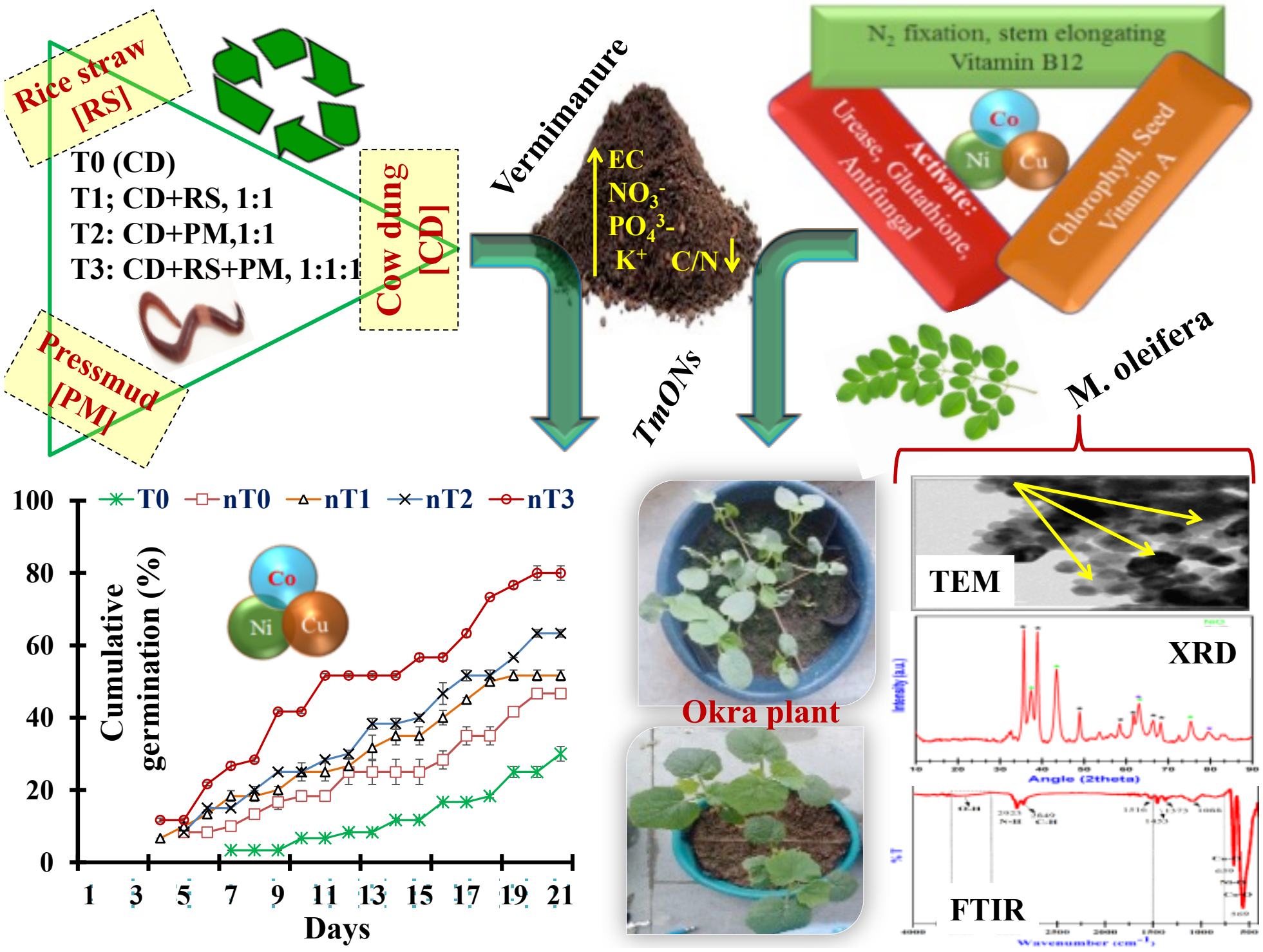

## Highlights

- Combined effect of Cu/Ni/Co oxide nanoparticles and agro-waste (RS and PM) based vermicompost as soil amendment.
- The synthesized nanoparticles (TmONs) were found to have CuO, NiO and CoO phases ranging from 13-54 nm.
- Increased N (26%), P (20%), and K (72%) in vermicompost by RS and PM.
- Toxicological risk by TmONs concentration has been evaluated.
- Significant plant growth was shown in nT3 (TmONs+ 1CD:1RS:1PM) compared to control.

## ABSTRACT

This study is a continuation of previous work, which highlights the nutrient enhancement by using rice straw (RS) and pressmud (PM) on vermicomposting. Herein, we demonstrate the significant impact of *Moringa oleifera* derived Cu/Ni/Co oxide nanoparticles (TmONs) in conjunction with these vermicompost on the growth performance of *Abelmoschus esculentus*. Vermicompost produced under various combinations (T0, cow dung (CD) only; T1, 1CD:1RS; T2, 1CD:1PM, and T3, 1CD:1RS:1PM) were further enriched by blending with biogenic nanoparticles. This strategic combination enhances the nutritional composition of the vermicompost, contributing to its overall effectiveness in promoting plant growth and health. Various analytical techniques, including FTIR, XRD, XPS, FESEM-EDX, TEM, and ICP-OES, were employed for comprehensive characterization. The synthesized TmONs with sizes ranging from 13 to 54 nm exhibited distinct CuO, NiO, and CoO phases. The vermicompost blended TmONs demonstrated significant improvements ($P < 0.05$) in seed germination (167%), coefficient velocity (67%), and vigour index (95%), while reducing the mean germination time by 41% for *A. esculentus* compared to the control group. The plant culture group nT3 (T3 + *TmONs*) showed the best growth performance. Furthermore, trace element concentrations in both soil and plant leaves were found to be below the maximum permissible limits set by WHO (1996). This investigation extends the understanding of the role played by these nanoparticles in fostering optimal conditions for plant growth and development as these micronutrients are essential components for several plant enzymes.

## 1. Introduction

Nanotechnology has emerged and gained the attention of researchers due to their numerous applications in various fields including, agriculture. Among the nanoparticles, metal oxide nanoparticles are still an emerging technology to encourage sustainable agriculture **(Bapat *et al.*, 2022 and Nguyen *et al.*, 2023a)**. The use of soil amendments in the form of metal oxide nanoparticles to alter soil properties which favours improved crop production is catching the attention of researchers. The main objectives of soil amendments are: (1) to provide nutrients, (2) to prevent insect and microbe-induced disease and (3) to increase nutrient availability in the soil **(Yan *et al.*, 2023)**. Because of their distinctive qualities, namely their small size with a high surface area to volume ratio, controlled release characteristics and sorption capabilities, these nanoparticles are regarded as suitable for a potential fertilizer formulation **(Nguyen *et al.*, 2023b)**. Several chemical routes for the synthesis of mixed metal oxide nanoparticles are present which utilize toxic chemicals which limits its biocompatibility and industrial-scale applicability. Thus, development of economical, scalable, biocompatible, and sustainable routes for the synthesis is required. In this context, biosynthesis of nanoparticles from plant extract is an attractive route. *M. oleifera* has been widely used in biogenic synthesis of nanoparticles as it contains a variety of phytoconstituents, which further are responsible for the development of the plant **(Thakur *et al.*, 2022 and Tufail *et al.*, 2022)**.

Simultaneously, the fast-growing worldwide population has led to significant increase in agricultural output leading to extensive use of agrochemicals and inorganic fertilizers. Unfortunately, excessive amounts of these are frequently lost through leaching which causes serious environmental pollution by contaminating groundwater and rivers **(Aniza *et al.*, 2023)**. Furthermore, the agricultural practices employed during the Green Revolution have raised serious concerns due to the production of agricultural waste that is indiscriminately burned or dumped in open areas leading to the creation of various pollutants and health consequences

**(Zhang *et al.*, 2019)**. However, two noteworthy agricultural wastes namely rice straw and pressmud (industrial sludge as compressed sugar), are among the most studied wastes which cannot be used as fertilizers as such alone and create pollution for both land and air **(Zong *et al.*, 2022)**. Due to the waxy nature and high sulphur content of pressmud, it can deteriorate the soil properties such as aeration, permeability, structure etc., which is likely to worsen with time. On the other hand, when burned incompletely, rice straw releases harmful gases like carbon mono-oxide, volatile organic compounds, and other carcinogenic aromatic compounds. Moreover, it also degrades the NPK content of the soil.

Therefore, the re-utilization of these agricultural wastes into value-added organic fertilizer by vermicomposting is regarded to be more valuable. It is a cost-effective, cleaner, and an eco-friendly choice that improves mineralization, humification, and generates bioconversion products **(Wang *et al.*, 2023)**. To enhance its quality, pest management and disease resistance properties, a nanotechnology-based approach is needed to be advantageous in agronomy. Nanoparticles enhance the nutrient-holding capacity of vermicompost and at the same time add antimicrobial properties, which can help suppress plant pathogens and reduce the reliance on chemical pesticides **(Keerthana *et al.*, 2021; Ruan *et al.*, 2023)**. Micronutrients like copper, cobalt, and nickel are essential for plants whose deficiency can lead to physical deformities in stem and leaves, decrease rate of photosynthesis, cause fungal infection, decrease yield, etc. **(Ahmad and Ashraf, 2011 and Ahmad *et al.*, 2022)**. Cu, Ni, and Co oxide nanoparticles have shown promising applications in environmental remediation, energy storage, and biomedical fields. Moreover, they are less expensive compared to other precursors, such as gold or silver which makes them economically viable for large scale production and commercialization. However, their combined potential in agriculture, particularly as a soil amendment, remains largely unexplored. **Deng *et al.* 2022** discovered crop-dependent responses of Cu nanoparticles in grain size, sugar and starch content, and element accumulation. Moreover, CuO-NPs increase iron concentrations in rice, implying that they can be used as nano-fertilizer. NiO-NPs significantly increased antioxidant enzymes and their activity in *N. arvensis* **(Chahardoli *et al.*, 2020)**. $Co_3O_4$-NPs have also shown positive effects on photosynthetic parameters of *Arabidopsis* plant **(Ogunyemi *et al.*, 2023)**. Apart from that, antibacterial effects of bivalent metal chelate (Co and Ni) against harmful soil bacteria such as *Staphylococcus aureus*, *S. faecalis* have been described by **Adewuyi *et al.* (2011)**. Collectively, the synergistic approach of these two metals holds promise for novel antibacterial drugs produced from vermicompost. Metal ions, such as $Cu^{2+}$, $Ni^{2+}$, and $Co^{2+}$, among others, comprise an important class of antimicrobial drugs with distinct active targets from most bacteriostatic polymers. So, combining these three metals as trimetallic oxide nanoparticles could open new avenues for sustainable crop management by supplying plant nutrients and protection **(Todaro *et al.*, 2023).**

The objective of this article was to shed light on the importance of metal ions of Cu/Ni/Co oxide nanoparticles as a plant micronutrient, its effects on plant growth and toxicity. We intended to raise awareness that Cu/Ni/Co oxide nanoparticles are a potential micronutrient of plants when used in low concentration. And further in-depth research is needed at molecular level to confirm this proposition. Overall, the combination of abundance, catalytic properties, versatility, cost-effectiveness, and environmental friendliness makes cobalt, copper, and nickel attractive choices for green nanomaterials.

This research endeavours to combine the individual strengths of *Moringa* extract, Cu/Ni/Co oxide nanoparticles and agro-waste (RC and PM) into a synergistic nano-organic amendment. This soil amendment is used in the right proportion to aid the growth profile of Okra (*A. esculentus*). Okra is an economically important vegetable with India as the leading producer. It offers a wealth of nutrients important to health including iron, amino acids, unsaturated fatty acids, vitamins, and minerals (**Manimaran *et al.*, 2022**). $SiO_2$, $Al_2O_3$, $TiO_2$ and Se nanoparticles have been used in recent studies for enhancing the phytochemicals and mitigating stress in okra **(Sonali *et al.*, 2023).**

This approach for the development of a novel nano-organic vermicompost outlining the impact of agricultural waste in the production of potential nutrient-rich soil amendment is newly introduced to the best of our information. This research employs variety of sophisticated analytical techniques, including field emission scanning electron microscopy with energy dispersive x-ray (FESEM-EDX), powder x-ray diffraction (XRD), ultra-performance liquid chromatography (UPLC-PDA), high-resolution transmission electron microscopy (HR-TEM), fourier-transform infrared spectroscopy (FTIR), x-ray photoelectron spectroscopy (XPS) and inductively coupled plasma optical emission spectroscopy (ICP-OES) to characterize trimetallic nanoparticles as well as to assess the maturity and qualitative evaluation of vermicompost and nano-vermicompost.

## 2. Results

### *2.1. TmONs synthesis and characterisation*

The UPLC with photodiode array (UPLC-PDA) shows the presence of such secondary metabolites namely, caffeic acid, ferulic acid, luteolin, *p*-coumaric acid, procatechuic acid, quercetin, rutin and vanillic acid in *Moringa oleifera* leaves extract. The trimetallic nano powder fabricated using *Moringa* extract was blackish in colour with about 60% yield in every repeated synthesis.

FTIR spectra of *TmONs* are shown in Figure. 1a. The broad absorptions near 3500 $cm^{-1}$ confirm the existence of O-H group stretching. The region of C-$CH_3$ and $CH_2$ stretching is present at 2923 and 2849 $cm^{-1}$ respectively. Small peak at 1715 $cm^{-1}$ is due to the presence of aliphatic ketone. The C=O stretching of amides I and II is shown by the peaks at 1516 and 1640 $cm^{-1}$. The C-H bending peaks are present at 1353 and 1373 $cm^{-1}$. Twin peaks at 1088 $cm^{-1}$ indicate presence of C-O linkage. The peaks corresponding to M-O bonds are present beyond 700 $cm^{-1}$. Particle size and defects are revealed by XRD patterns, while the atomic distribution within the unit cell is shown by relative peak intensities. Figure. 1b depicts XRD patterns of the synthesized *TmONs*. The diffraction peaks at 32.54°, 35.56°, 38.77°, 48.74°, 53.53°, 58.37°, 61.56°, 66.29° and 68.17° can be indexed to (110), (002), (111), (-020), (020), (202), (113), (022), (311) planes of monoclinic phase of CuO (Ref. 00-041-0254), respectively **(Che *et al.*, 2022)**. Moreover, the well-defined diffraction peaks at 37.18°, 43.12°, 62.88°, 75.24° and 79.23° are ascribed to (111), (200), (220), (311) and (222) planes of rhombohedral NiO (Ref. 00-044-1159) **(Choudhury *et al.*, 2018)**. Similar XRD patterns for the planes of cubic CoO (Ref. 00-002-1217) can be assumed as the mixing of both oxides.

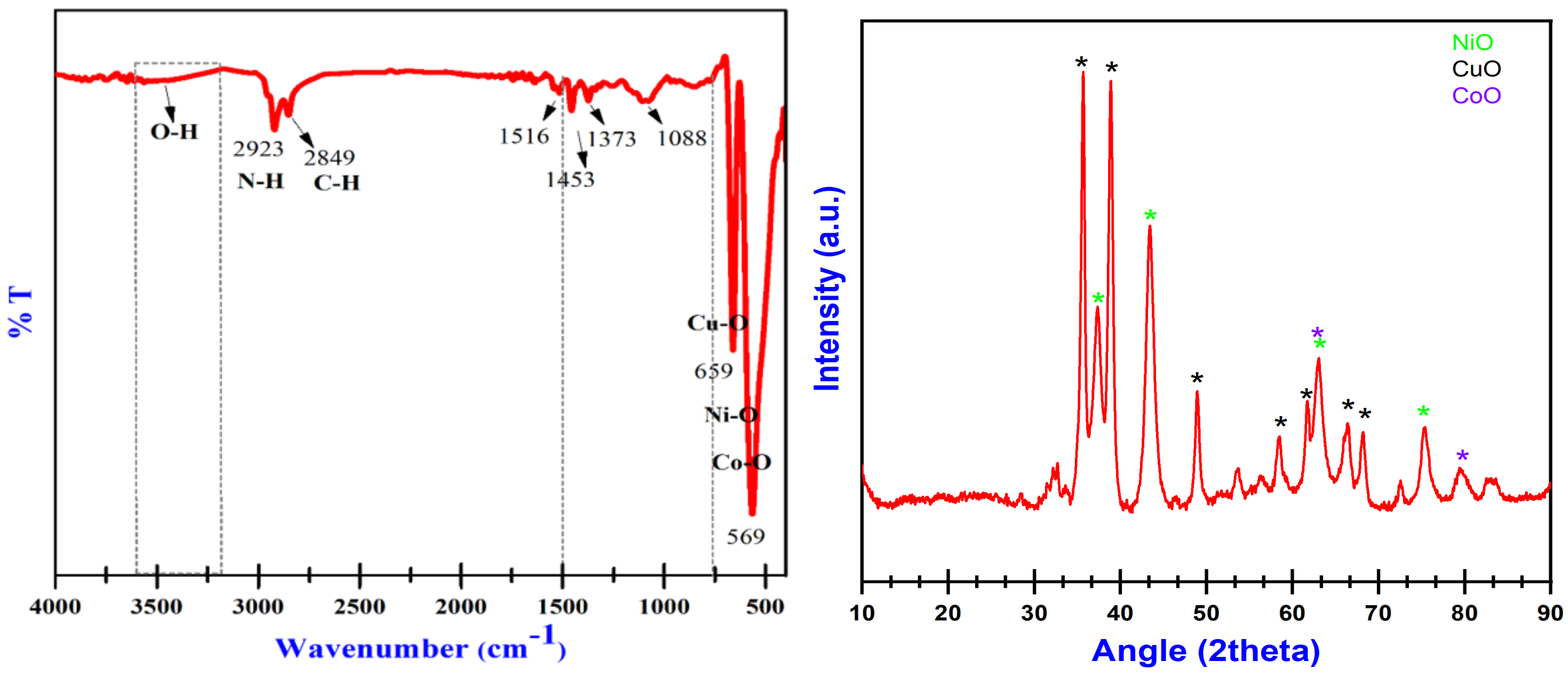

**Figure. 1.** FTIR spectrum (a) and XRD patterns (b) of TmNOs.

The chemical states of *TmONs* and their specific orbitals were also confirmed by XPS measurements. Figure. 2a shows the survey scan, which clearly indicates the existence of Cu, Ni, Co, C, N, and O with some auger peaks corresponding to metal oxides. The core-level spectra were fitted with Gaussian function to separate the overlapping contribution that may appear due to presence of the element in different environment. The background that results from inelastic scattering of electrons was subtracted by employing a Shirley background correction. The characteristic peaks for binding energies of N 1s, O1s and C 1s are present at 399, 531.9 and 284 eV respectively. Four peaks in the Cu2p region were seen in the high-resolution XPS spectra of *TmONs* shown in Figure. 2b. Cu2p $_{3/2}$ and Cu2p $_{1/2}$ were both represented by the peaks that were detected at 933.6 and 953.38 eV, respectively **(Mani *et al.*, 2023)**. $Cu^{2+}$ ion satellite peaks are located at 942.1 and 962.0 eV which is a feature of partially filled d-block ($3d^9$) $Cu^{2+}$. It ruled out the occurrence of $Cu^+$ ($Cu_2O$) and proved the presence of $Cu^{2+}$ ions (CuO) **(Rajendran *et al.*, 2021)**. Similarly, the high-resolution XPS spectra of Co2p may also be fitted into four Gaussian peaks. The low-binding energy (Co2p $_{3/2}$) and high-binding energy (Co2p 1/2) at 780.2 and 795.7 eV accompanied with the shake-up satellites at 783.6 eV and 802.7 eV match well with $Co^{2+}$, respectively **(Rozina *et al.*, 2022)**. In Ni2p spectra, the major peaks located at 861.26 eV, 855.9 eV and 854.36 **(Singh *et al.*, 2022)** are distinctive NiO peaks (Figure. 2d). Based on the area under the peaks, the atomic percent of the elements of nanocomposites are 2.83%, 2.17% and 1% for Cu, Ni and Co respectively, which is nearly consistent with the ratio initially taken.

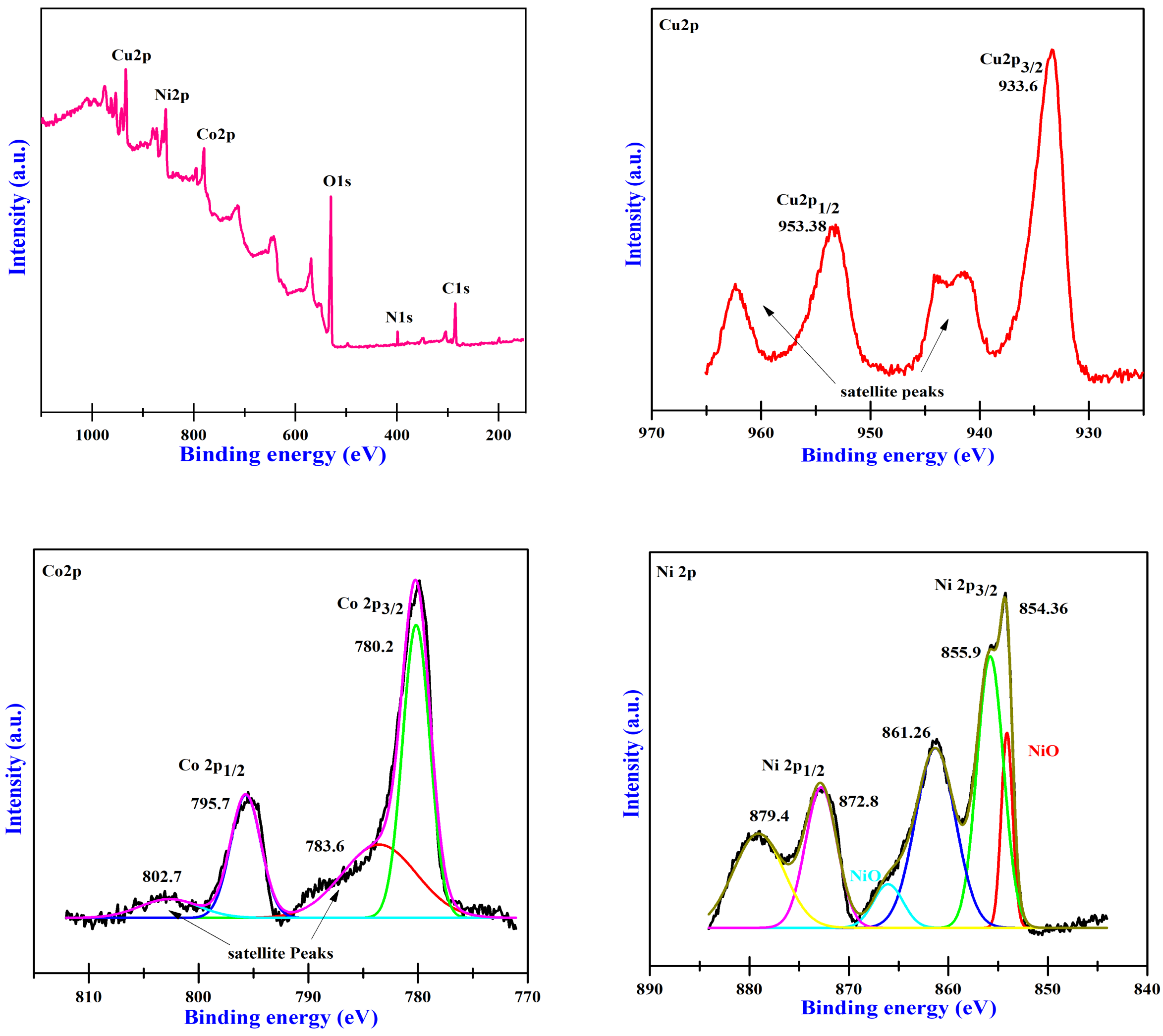

**Figure. 2.** XPS survey scan, Cu 2p spectra, Ni 2p spectra and Co2p spectra

The EDAX analysis and mapping also confirmed the TmONs contain Cu, Ni, Co, C, N, and O elements, as shown in Figures S1. The morphological shape, size and crystalline phase of the nanocomposites were analysed by TEM at an acceleration voltage of 200 kV. The nanocomposites were submerged in ethanol at room temperature for 10 minutes before being ultrasonically dispersed, and the samples were placed on copper grids coated with amorphous carbon sheets. Figure. 3 depicts the various morphologies of each metal oxide, such as hexagonal, spherical, and nearly spherical. As expected from green synthesis, there is some particle agglomeration. The substance was found to include nanoparticles ranging in size from 14 to 54 nm, with a mean size of 29 nm. The d-spacing, which denotes the minimum distance between two successive planes in a lattice was calculated by using ImageJ software and found to be 0.266 nm and 0.206 nm, which is quite consistent with the d-spacing obtained through XRD. In ImageJ, after setting the scale of the HR-TEM image to nm unit, FFT (Fast Fourier Transformation) and inverse FFT image filtering process were performed. This process resulted in lattice fringes and the distance between two parallel lines were calculated manually as d spacing. Furthermore, the SAED pattern shows diffraction rings with particle spots indicating the polycrystalline nature of the *TmONs*.

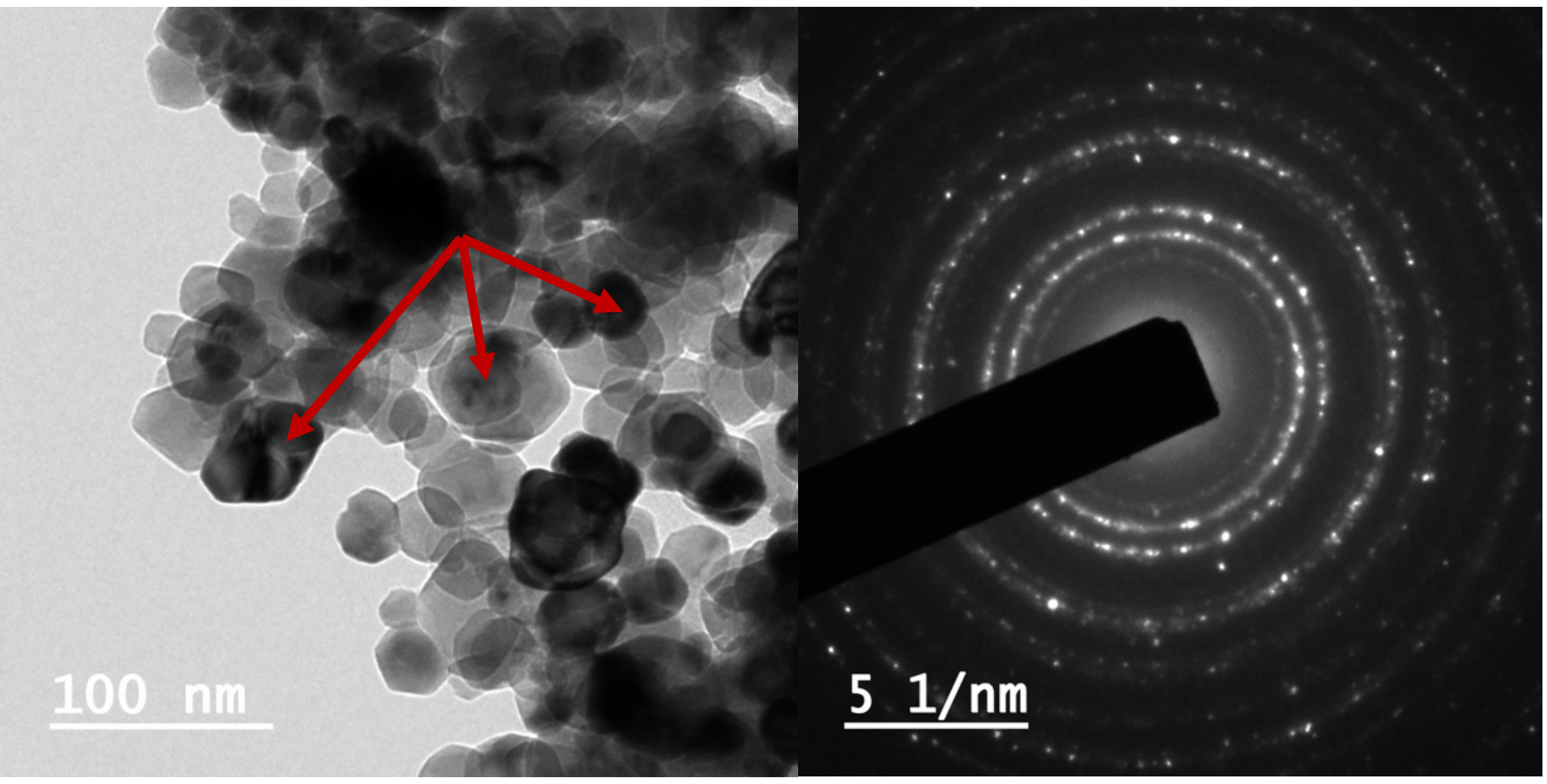

**Figure. 3.** TEM image along with SAED pattern of TmONs

*2.2. Physico-chemical characteristics, nutrient, and surface analysis of vermicompost*

Table 1 represents a typical nutritional status in terms of C/N, electrical conductivity (EC), pH, and the percentages of NPK used for vermicomposting, whereas Table S4 illustrates the variations in the initial raw materials used as feedstock and the vermicompost produced. As shown in Figure. 4, after 110 days of vermiculture, there was a statistically significant ($P < 0.05$) rise in the percentage of NPK macronutrients. For all the treated groups, the desired EC of <8 dS $m^{-1}$ was observed for the final product, and it was slightly higher in case of T3 (CD + PM) feedstock group in comparison to others. The pH of the developed vermicompost was observed to be within an optimum range of 7.1-8.1. The pH in the treated groups varied between 7.1 and 7.6 and stabilized at a value close to neutral on day 90.

**Table 1**

Initial chemical characteristics of the cow dung (CD), rice straw (RS) press mud (PM)and Soil (Mean ± SE, n=3)[a].

| **Parameters** | **Cow dung** [a] | **Rice straw** [a] | **Pressmud** [a] | **Soil**[a] |
|---|---|---|---|---|
| pH | 8.34 ± 0.36[a] | 8.5 ± 0.31[a] | 7.03 ± 0.10[b] | 9.03 ± 0.57[c] |
| EC ($dSm^{-1}$) | 9.89 ± 0.59[b] | 3.78 ± 0.10[a] | 4.77 ± 0.36[a] | 4.77 ± 0.36[a] |
| TOC (%) | 38.74 ± 2.44[c] | 7.50 ± 0.78[a] | 20.00 ± 1.93[b] | 41.51 ± 4.30[c] |
| N (%) | 1.18 ± 0.08[b] | 0.72 ± 0.10[a] | 1.98 ± 0.23[c] | 0.75 ± 0.1[a] |
| P (%) | 0.13 ± 0.02[b] | 0.05± 0.02[a] | 0.33 ± 0.05[a] | 0.03 ± 0.01[a] |
| K (%) | 1.28 ± 0.17[c] | 0.59 ± 0.15[b] | 0.51 ± 0.05[b] | 0.07 ± 0.02[a] |
| C/N | 33 ± 1.93[b] | 11 ± 2.36[a] | 10 ± 1.71[a] | 56 ± 5[c] |

A substantial change in N% was reported in all treated feedstock groups as follows: T0 (1.39 ± 0.07) <T2 (1.50 ± 0.12) <T1 (1.57 ± 0.06) <T3 (1.82 ± 0.11). However, the N content of the vermicompost from the T3 fed group (CD + RS + PM) was higher compared to the T2 (CD + PM), T1 (CD + RS) and T0 (Control) fed feedstock groups, respectively by 21%, 16% and 31%. As a result, it was observed that the N content in all feedstock groups had increased overall by approximately 62% as compared to the raw materials used initially (Figure.4).

All treated groups had a higher proportion of P macronutrient after complete process of vermicomposting (110 days) than the control (T0) feedstock group. P% changed significantly (P<0.05) throughout the procedure. Among all, the P content of the T3 (CD +PM + RS) vermicompost group was comparatively higher by 14.29%, 41.18% and 60% than that of the T2 (CD + PM), T1 (CD + RS), and T0 (Control) respectively (Figure.5).

The potassium (K) concentration in the blended (RS+PM) feedstock groups was discovered to be higher than the control (T0) feedstock group and the original raw materials used for vermicomposting. The mature vermicompost feedstock T2 (CD + RS) had the highest K% followed by treatments T3 (CD + RS + PM) and T0 (Control). Quantitatively it was 34.8%, 19.5%, and 3.5% higher than T0 (Control), T1, and T3 fed feedstock groups, respectively.

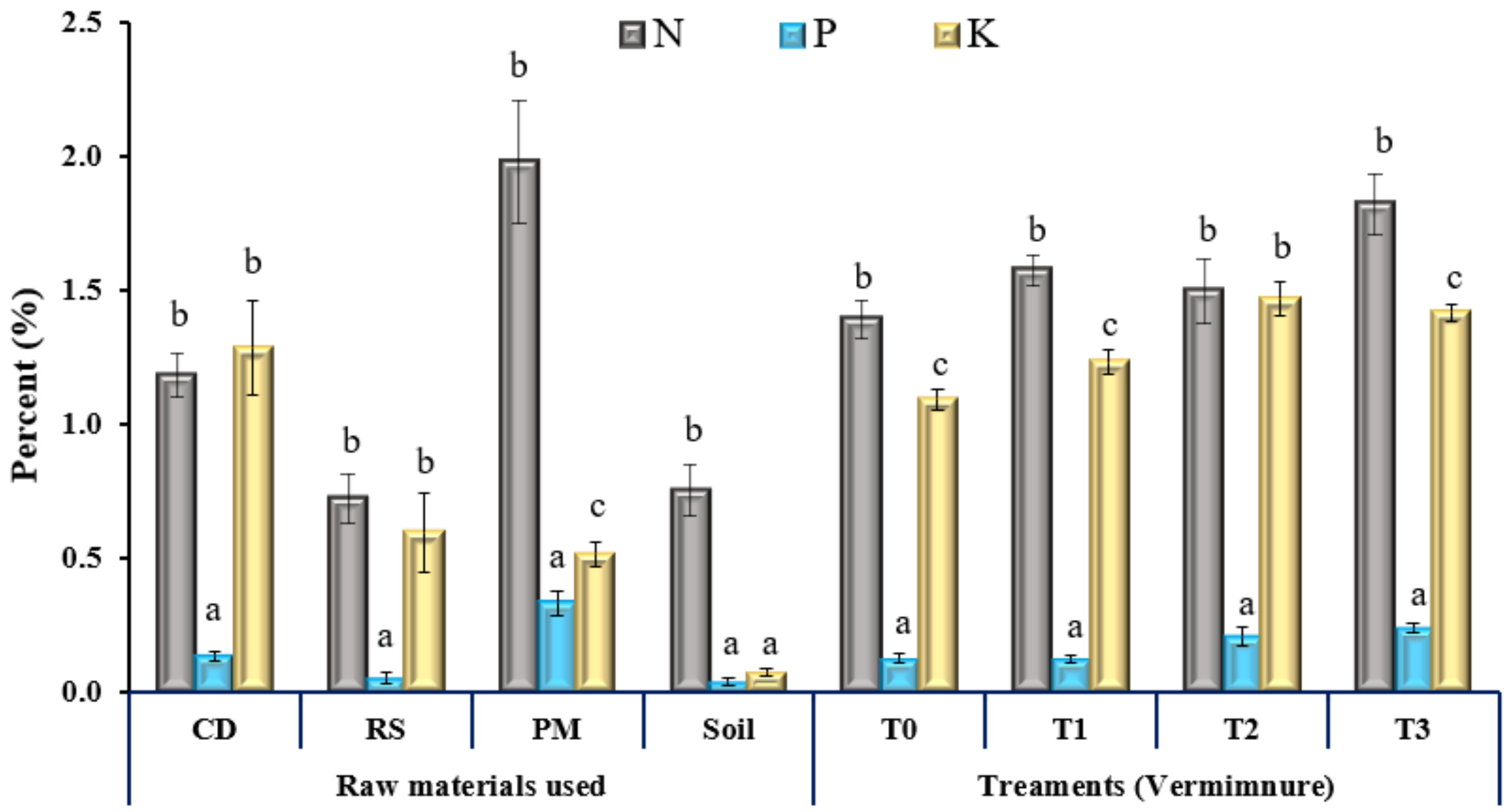

**Figure. 4.** Comparision of N, P and K concentration in raw material used and produced vermicompost. Bars with different letters are significantly different ($P < 0.05$).

The surface structure and elemental makeup of all produced vermicompost from various feedstock combo were determined using the FESEM-EDX technology. Vermicompost were found to be more disaggregated than controls, but no significant changes were seen after TmONs treatment (see Figures S2 and S3). When vermicompost samples from different feedstocks were compared, (T3) had the highest proportion of rough matrix surface, followed by (T2), (T1), and (T0).

### *2.3. Toxicological risk assessment of heavy metals on earthworms, soil, and plant*

Earthworms are always employed as the model organisms for toxicological risk assessments since they are also considered to be monitors and natural indicators of soil health (**Huang *et al.*, 2023).** The earthworms were exposed to different concentrations (50, 100, 200, 300, and 500 ppm) of *TmONs* and it was found that earthworm under 500 ppm and 200 ppm dose of *TmONs* showed the highest mortality (100% and 90%) followed by 300 ppm (60%). Maximum survival was found in 50 ppm (90%) and 100 ppm (70%) treated groups. Therefore, in our study, 100 ppm *TmONs* concentration was used with soil (vermi + soil) for the evaluation of plant growth performance.

Acid digestion, also known as wet digestion, reduces interferences by breaking down the sample matrix, making it efficient for organic compounds **(Uddin et al., 2016)**. According to ICP-OES, all metals (Cu, Ni and Co) were found to be under permissible limits in soil (Cu: 0.028-0.162 ppm; Ni: 0.008-0.162 ppm; Co: below detection limit) as well as Okra leaves (Cu: 0.070-1.408 ppm; Ni: 0.188-1.203 ppm; Co: below detection limit) as shown in Table 2. All samples of soil and plant leaves were below the permissible limits set by **WHO (1996)**.

**Table 2**

ICP-OES analysis of soil samples after growth of *A. esculentus*

| Treatments* | Sample | Cu (ppm) | Ni (ppm) | Co (ppm) |
|---|---|---|---|---|
| T0 | Soil | 0.028 | 0.008 | -0.166 |
| | Plant | 1.488 | 0.460 | -0.097 |
| nT0 | Soil | 0.162 | 0.066 | -0.150 |
| | Plant | 0.070 | 0.328 | -0.348 |
| nT1 | Soil | 0.019 | 0.077 | -0.089 |
| | Plant | 0.078 | 0.188 | 0.891 |
| nT2 | Soil | 0.060 | 0.162 | -0.430 |
| | Plant | 0.991 | 0.953 | -1.254 |
| nT3 | Soil | 0.089 | 0.115 | -0.229 |
| | Plant | 1.408 | 1.203 | -2.223 |
| Permissible limits (soil) WHO (1996) | | 30 | 80 | 50$ |
| Permissible limits (plant) WHO (1996) | | 10 | 10 | 0.2-0.5# |

*T0 (without *TmONs*), positive control as nT0 (T0+*TmONs*), nT1 (T1+*TmONs*), nT2(T2+*TmONs*) and nT3 (T3+*TmONs*). #maximum tolerable limit **(Mahey *et al.*, 2020)**; $**Lowel and Joel 2020**.

*2.4. Growth performance of A. esculentus*

The initiation of seed germination commenced on the fourth day after sowing the seeds in soil amended with vermicompost and TmONs. The recorded germination events for seeds in the T0 group were documented on the seventh day. In contrast, in groups treated with TmONs (nT0-

nT3), the first germination instances were observed on the fourth day post-sowing, as illustrated in Figure 5a. The T0 group exhibited a modest germination rate, with only 7 to 12% of plants germinating within the 10 to 15-day timeframe. In contrast, the groups treated with TmONs (nT0-nT3) displayed enhanced germination rates, ranging from 18 to 25% (nT0), 25 to 35% (nT1), and 25 to 40% (nT3). As the observation period extended to 21 days post-sowing, the germination percentages increased to 30% in the T0 group, 47% in nT0, 52% in nT1, 63% in nT2, and 80% in nT3, as depicted in Figure. 5b. These findings unequivocally demonstrate that the incorporation of TmONs into the vermicompost significantly augmented both the germination rates and seedling growth. The enhanced performance observed across the nanoparticle-treated groups highlights the positive impact of TmONs on seed germination and early seedling development.

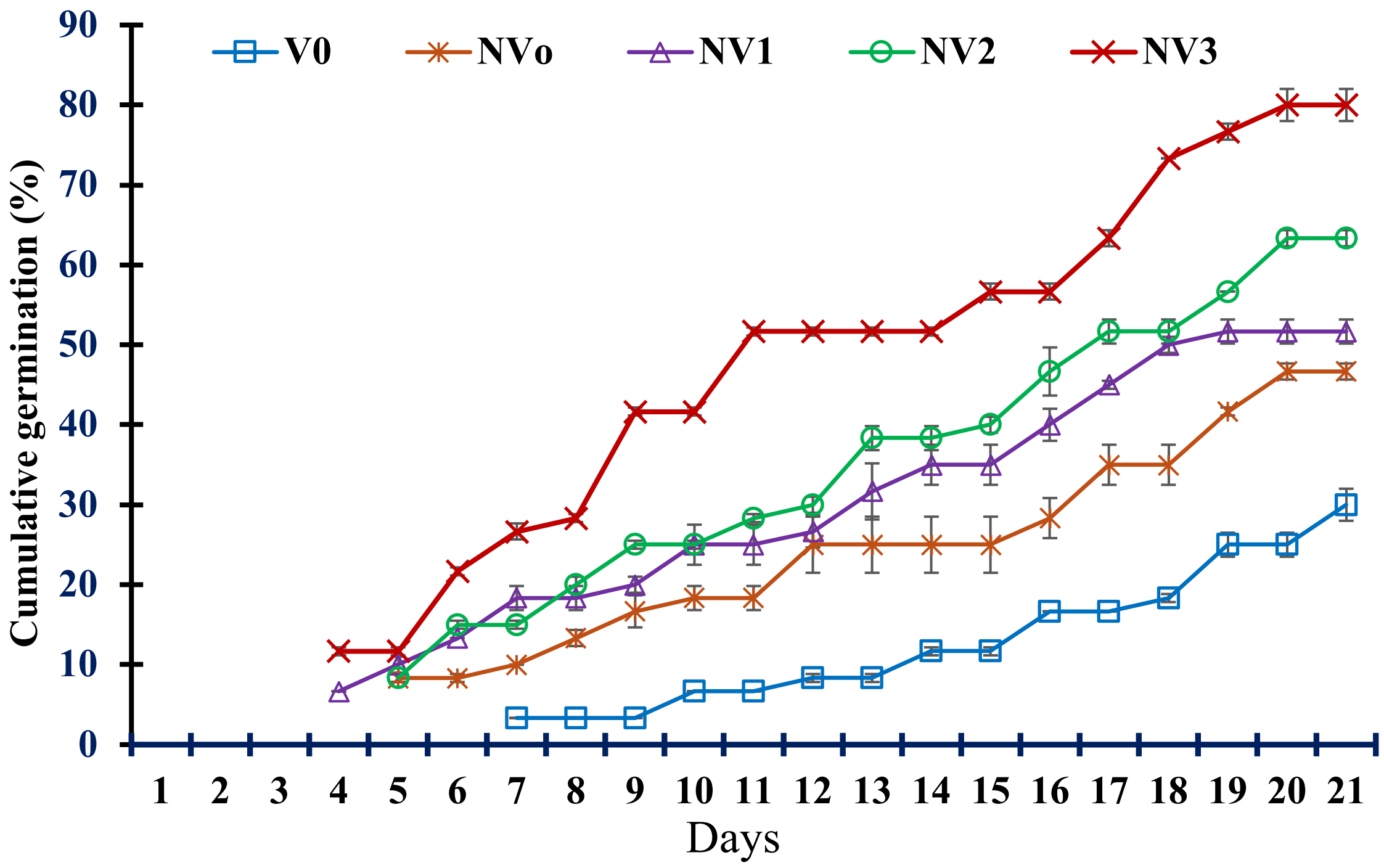

a)

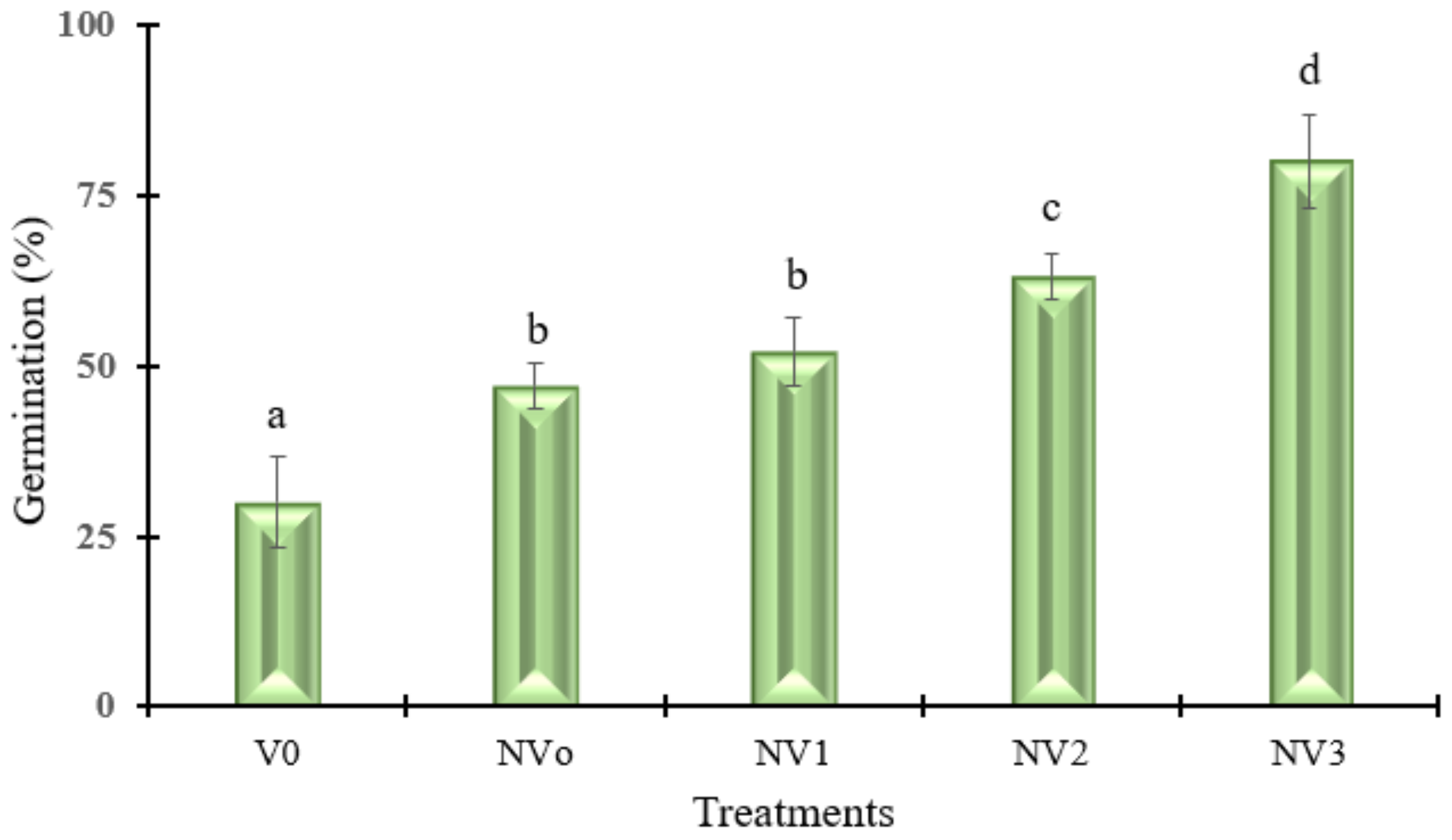

**b)**

**Figure. 5.** Cumulative germination (a), and Germination percent (b). Bars with different letters are significantly different ($P < 0.05$).

The MGT serves as a crucial parameter, representing the average duration required for the maximum germination of seedlings. In comparison to the negative control group, a significant ($P < 0.05$) reduction in MGT was observed across all groups treated with TmONs blended with vermicompost, as depicted in Figure 6a. Remarkably, the lowest MGT value was recorded for the nT3 group (10.1± 0.9), signifying a rapid and efficient germination process. The order of decreasing MGT values among the treated groups, as illustrated in Figure 6a, was T0 > nT0 > nT1 > nT2 > nT3. Quantitatively, the MGT values were 27.5%, 51%, 41%, and 63% lower for nT0, nT1, nT2, and nT3 groups, respectively, when compared to the control (T0) group. These findings highlight the significant improvement in germination efficiency facilitated by the incorporation of trimetallic oxide nanoparticles into the vermicompost. The substantial reduction in MGT across the treated groups underscores the positive impact of TmONs on accelerating and optimizing the germination process, thereby promoting robust seedling emergence.

The CVG serves as an important metric in assessing the speed and uniformity of seed germination. In our study, the highest and lowest CVG values were observed for the nT3 group (10 ± 1.0) and the T0 control group (6 ± 0.5), respectively, as illustrated in Figure. 6a. Additionally, the CVG for the nT3 group exhibited an approximate increase of 67%, 25%, 5%, and 18% over the T0, nT0, nT1, and nT2 groups, respectively. Despite these variations, the values for CVG did not demonstrate any statistically significant differences between the vermicompost mixed with TmONs.

Seed vigor, a crucial indicator of seed quality with a direct impact on the potential for plant emergence (Wen et al., 2018), is assessed through the SVI. The SVI is calculated by multiplying the germination percentage and seedling length (mm) (Qijuan et al., 2017). Across all nT0-nT3 groups, the SVI values were consistently higher than the T0 control group (Figure. 6b). However, when comparing SVI values among the treated groups, the order was as follows: T0 (840 ± 30) < nT1 (1118 ± 78) < nT0 (1269 ± 23) < nT2 (1355 ± 95) < nT3 (1640 ± 40). Furthermore, the SVI in the nT3 treatment was notably 29.2%, 46.7%, and 21.03% higher than the nT0, nT1, and nT2 groups, respectively. These results collectively emphasize the positive influence of trimetallic oxide nanoparticles on both the speed and vigor of seed germination, highlighting the potential for improved plant emergence and growth in nanoparticle-treated groups.

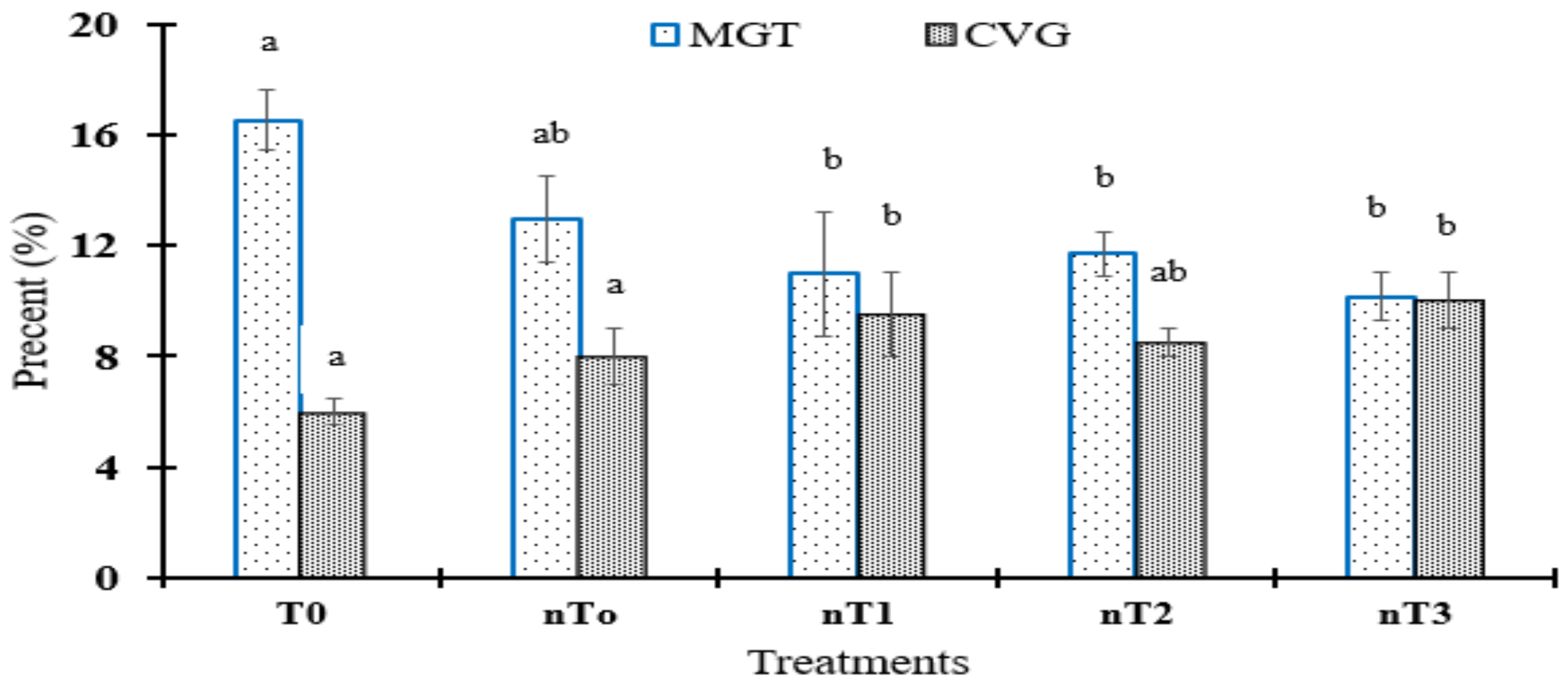

a)

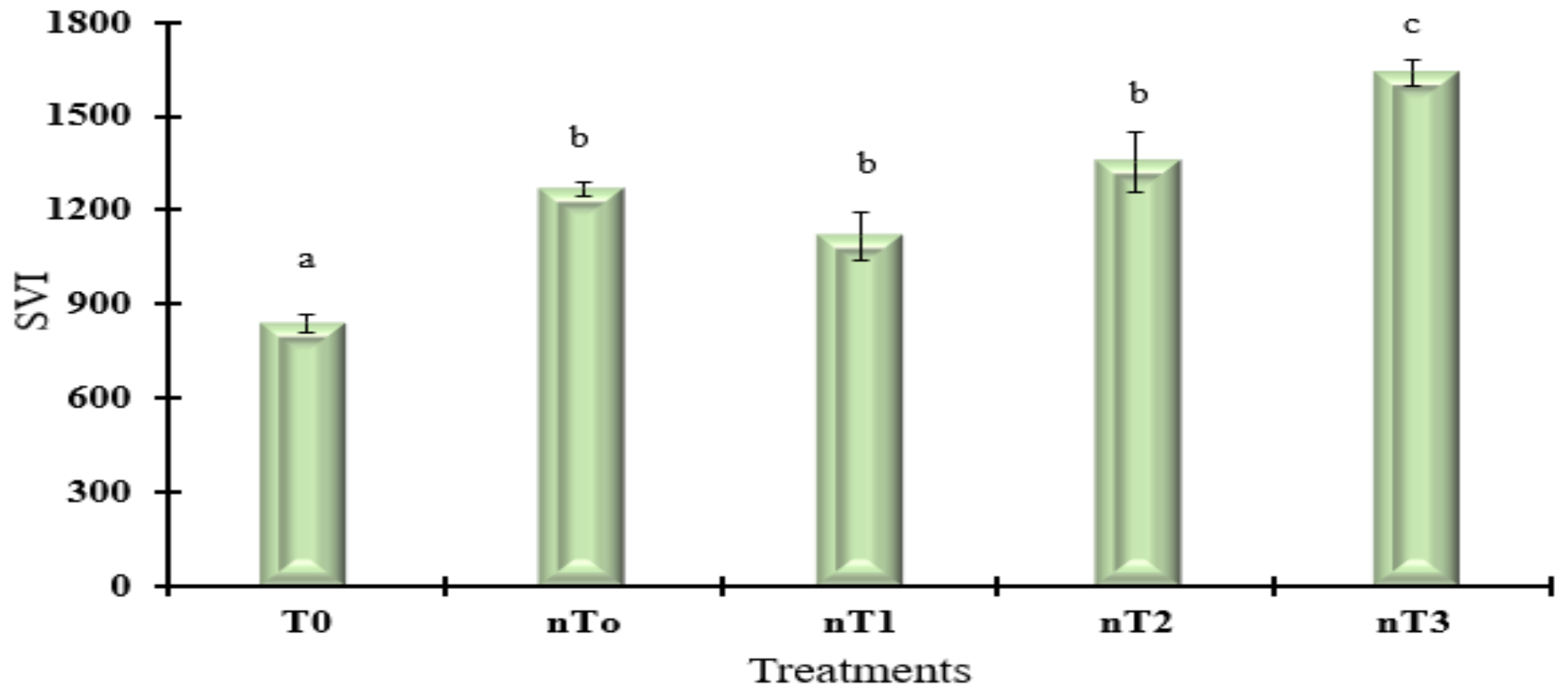

b)

**Figure. 6.** Mean Germination Time (MGT), Coefficient Velocity of Germination (CVG) (a), and Seed Vigour Index (SVI) (b). Bars with different letters are significantly different ($P < 0.05$).

## 3. Discussion

Phytoconstituents of *M. oleifera* possess the ability to serve both as reducing and capping agent for the synthesis of TmONs due to the presence of various phenolic acids and polyphenols. Researchers have developed a number of variants for standard solution combustion method. For instance, some prefer heating the solution of precursor and fuel (urea, aminophenol etc.) on hot plate while some prefer to form a paste using mortar and pestle and further place it in muffle furnace. Using plant extract as an eco-friendly fuel alternative in the synthesis further adds to the biocompatibility and removes toxicity. A similar method has been used by **Maruthupandy *et al.*, 2017**, for the preparation of CuO and ZnO nanoparticles for their potential application in sensing contamination caused by $Li^+$ and $Ag^+$ ions in marine water samples. To prepare TmONs, the ratio of salts were taken following the Irving-William series ($Co^{2+}$<$Ni^{2+}$<$Cu^{2+}$) as these metals mimic the naturally occurring metalloenzymes and binds in order of affinity according to the series. Hence, 3g copper precursor was mixed with 2g nickel salt and 1g cobalt salt, that also takes care of the particular metal concentration and toxicity.

EC, pH, and C/N ratio are considered maturity parameters of vermicompost. EC is widely used to measure manure salinity, which may be reduced as a result of earthworm aided decomposition of organic matter as earthworms' bodies biologically accumulate certain minerals, which has led to a decrease in the soil's mineral content **(Devi and Khwairakpam, 2023)**. EC arises as a result of breakdown of organic matter in soil and freely available salts/ions created during earthworm metabolism. Therefore, the higher value of EC in all blended feedstock group (except T0) might be due to a higher concentration of salts/ions **(Alavi *et al.*, 2017)**. A lower C/N ratio may suggest an excess of nitrogen, while a higher ratio may indicate an excess of carbon **(Dume *et al.*, 2023)**. It's important to note that the C/N ratio of vermicompost may vary based on the initial mix of materials used in the composting process. As a result, the C/N ratio of the T3 (CD+RS+PM) fed group was shown to be considerably ($P < 0.05$) lower than the other feedstock groups. A vermicompost with a C/N between 12 and 25, and preferably less than 15, is considered a good indication of mature vermicompost **(Jiménez and Garcia, 1989)**. The pH variation in the range 7.1-8.1 was ascribed to a surge in nitrogenous waste excretion during vermicomposting, which had the ability to bring back the pH level to neutral **(Hujuri *et al.*, 2023)**.

It was shown that earthworm castings contain more NPK than substrate materials and these elements are in the forms that plants can readily use which in turn increases the soil's NPK content and lowers the C/N ratio **(Devi and Khwairakpam, 2023)**. Our study showed increased NPK after vermicomposting, one major factor contributing to this increase in P is the activity of gut enzymes which are released in different regions of the earthworms' gut. As for the plant, the photosynthetic process relies on P content, which is also necessary for soil nutrient uptake. Nitrogen content can alter the chlorophyll production and hence, photosynthesis and biomass accumulation **(Kumar *et al.*, 2023; Kumari *et al.*, 2023)**. However, when compared to chemical fertilizer, vermicompost contains less NPK and more organic compounds like, polysaccharides, proteins and lipids which boost the microbial colony **(Ali *et al.*, 2024)**. Due to the eco-friendly nature, vermicompost feeding could greatly influence crop quality management by accessing the problems associated with chemical fertilizers **(Ruan *et al.*, 2023)**. The more disaggregated vermicompost as compared to controls could be due to the action of bacteria and enzymes produced by *E. fetida*, which was further enhanced by adding nanoparticles in nT3. High surface area, tunable porosity and strong cation exchange interactions could have resulted in massive holes, fragmentation, and porosity allowing good absorption, aeration, and drainage **(Srivastava *et al.*, 2023)**.

The synergistic effect of TmONs, pressmud, and rice straw may explain the increased germination, CVG, SVI, and lower MGT. A lower MGT denotes the probability of seed emergence **(Khajeh *et al.*, 2009)**. The fastest growth of the population of *okra* seed for nT3 is attributed to the statistically lowest MGT value. The ability of nanomaterials to penetrate the seed coat and speed up embryogenesis by enzyme activity may have led to seed dormancy break down and could have accelerated the germination process **(Younes *et al.*, 2020)**. The values for CVG showed no significant relation between the vermicompost mixed with TmONs. These co-efficient values vary from 0 to 100 with high values towards 100 suggesting rapid germination and low values near 0 indicating very sluggish or even blocked germination. Overall, this co-efficient is used to calculate the rate and spread of germination, i.e., it is a pure rate measurement devoid of germinability interference **(Talská *et al.*, 2020)**. In the field or during storage, it is equally important to analyse the seed for quality characteristics in addition to germination and other viability tests. Nowadays, seed companies employs seed vigour to determine seed quality. High seed vigour obtained in our study could be attributed to the potency of soil-amendment in enhancing growth and productivity **(Seyyed *et al.*, 2015)**. Furthermore, several works have also been reported for the growth of plants using different metal oxide nanoparticles (Table 3).

Germination involves complex biochemical events for instance, functional activation of plant hormones, modification of lipids, degradation of storage proteins, and cell walls, biosynthesis of NADPH, ATP, DNA and RNA **(Wu *et al.*, 2020)**. Nanoparticles can interact with soil in various ways for instance, interaction with root exudates, humic acid, organic reducing substances and bacterial colony. It was found that entry of NPs in root cells is enhanced by interaction with organic acids and proteins inside plant tissue **(Chung *et al.*, 2019)**. It could be suggested that the ions $Cu^{2+}$/$Ni^{2+}$/$Co^{2+}$ were absorbed on the surface of the seeds and gradually released to show their effect during the test period. It could be due to the interaction with organic chelating agents present in the amended soil which reduced the movement, uptake and hence toxicity of these metals. Root exudates can also modify the NPs surface chemistry and alter their dissolution, and toxicity **(Shang *et al.*, 2019)**. **Xie *et al.*, (2023)** found an increase in the levels of various lipids in plumule and cotyledon of mung bean seed during initial germination.

**Table 3**

Some previous studies of growth parameters (MGT, CVG, %G and SVI)

| Nanoparticles/Dosage | Plant | Parameters | | | | **Reference** |
|---|---|---|---|---|---|---|
| | | MGT | CVG | %G | SVI | |
| ZnO (10-200 mg/L) | *Zea mays* | - | 2932.9 | - | - | Itroutwar *et al.*,2020 |
| n$SiO_2$ (2-14 g/L) | Tomato | 3.98% | 507.82% | 22.16% | 22.15% | Siddiqui and Al-Whaibi, 2014 |
| CuO (0.1-10 mg/L) | Wheat | 1.92±0.054 | 50.6±1.46 | 100% | - | Ibrahim *et al.*, 2022 |
| Fe ((25-250 mg/L)) | Soyabean seeds | - | - | 99.34% | - | Hoe *et al.*, 2018 |
| ZnO (25-500 mg/L) | | | | 94.74% | | |
| Cu (5-50 mg/L) | | | | 96.71% | | |
| Co (0.05-2.5 mg/L) | | | | 95.36% | | |
| MgO (10-100 mg/L) | *Vigna radiata* | - | - | 100% | 1340±10 | Vijai *et al.*, 2020 |
| ZnO (5-20 mg/L) | Wheat | - | - | 100% | 94% | Rai-kalal and Jajoo, 2021 |
| ZnO (750-1250 mg/kg) | *Capsicum annum* | - | - | 73% | 1177 | Kumar *et al.*, 2019 |
| $TiO_2$ (750-1250 mg/kg) | | | | 67% | 993 | |
| Ag (750-1250 mg/kg) | | | | 69% | 1075 | |
| **CuO/NiO/CoO (100ppm)** | ***Abelmoschus esculentus*** | **10.10±0.9** | **10±1.0** | **80%** | **1640±40** | **Present work** |

MGT: Mean germination time, CVG: Coefficient velocity of germination, %G: Percent germination and SVI: Seed vigour index.

Furthermore, these elements (Cu/Ni/Co) have been discovered to be a component of several enzymes involved in photosynthesis, nitrogen metabolism, electron transport, plant defence mechanisms, secondary metabolites, stress tolerance, antibacterial and antifungal capabilities, and so on **(Azarin *et al.*, 2022; Mariyam *et al.* 2023 and Vindhya and Kavita, 2023)**. The combined applications of *TmONs* optimise the concentration of individual components, which serve as an effective way to tolerate Cu, Ni and Co stress and toxicity since previous studies have confirmed that the size and concentration of nanoparticles plays a vital role in the uptake and relocation of the metal ions released from these oxides. **Ahmed *et al.*, (2018)** mentioned a concept of the oxidative window for germination, which is a critical range of the ROS level at which the occurrence of the cellular events associated with germination is optimal. ICP-OES results obtained well justify the above concept as all the concentrations within leaves and soil are under optimum ppm levels. Higher concentration of NiO, 250 mg/L and 500 mg/L caused toxicity in Chinese cabbage seedling while lower concentration of 50 mg/L had no such effects **(Chung *et al.*, 2019)**.

**Paulraj *et al.* 2022**, have studied the production of copper oxide nanoparticles (CuO NPs) using *Eudrilus eugeniae* vermi wash identifying its practicality in the germination mechanism of *Vigna radiata* seeds. Their results show that CuO NPs have a favourable effect on seed germination and their lower concentrations can be used as fungicides while high dose can alter the levels of acidic, basic amino acids and total protein **(Haider *et al.*, 2023 and Hasanin *et al.*, 2023)**. In our study, we had also seen similar antifungal effect for the treated plants with respect to control. These NPs may also get coated by root exudates or other organic moity present in the soil which can lead to surface modifications and hence reduce the toxicity of metals. This fact is further justified by **Deng *et al.* (2022)** where a coating of citric acid reduced the toxicity of CuO NPs in soyabean.

Furthermore, nickel is also an essential micronutrient for seed germination to senescence. It is a major component of urease enzyme, which hydrolyses urea into ammonia and carbamic acid, a source of nitrogen for plants. It increases the solubility of nitrogen and phosphorus in the soil making these nutrients more available to plants **(Zhou *et al.*, 2023)**. Along with this, nickel serves in nitrogen fixation along with cobalt and vitamin B12. It also helps in regulating the plant's stress response through the generation of reactive oxygen species in plants, thereby triggers the stimulation of several signalling pathways and the expression of stress-responsive genes. Germination was considerably enhanced in *Vigna mungo* plants when NiO-NPs were supplemented in a dose-dependent manner **(Rahman *et al.*, 2021)**. Furthermore, they have also been found to enhance the length of the roots and shoots and hence the overall biomass **(Chahardoli *et al.*, 2020)**.

Following copper and nickel, cobalt's relatively lower content may have aided in better nodulation and, as a result, improved growth, and yield in our study. In fact, specific mechanism of accumulation, translocation and specific transporter is still unknown to a great extent for cobalt. It is considered a non-essential element in plants, but recent studies have marked the vital role of Co in a number of physiological processes involving synthesis of photosynthetic pigments although it is still a neglected from a farmer's point of view because of lack of knowledge **(Banerjee *et al.*, 2021).** It has been suggested to govern the general vegetative and reproductive aspects of various crops like wheat. Instilling drought and stress resistance within the plant system is another essential function of CO, since it plays a regulatory role in reducing transpirational rate through water use regulation **(Faisal *et al.* 2023)**. The study by **Brengi *et al.* (2022)** provides valuable insight about the role of Co in redox homeostasis over melatonin in improving yield under salt stress conditions in cucumber. In a study, colloidal solutions of Co and Cu were added in Murashige and Skoog nutrient medium leading to increment in microplant height and growth index **(Talankova *et al.*, 2016)**. Moreover, the increased root hair formation in all the treated plants (Figure. 7a) could be due to the effect of Co which led to increased absorption of water and nutrients, further boosting growth. However, cobalt oxide NPs may also improve plant nutrient uptake by chelating with other metal ions and making them more available to plants. Furthermore, an ATP-binding cassette (ABC) transporter aids in the transportation of Co and Ni into and out of the chloroplast **(Li *et al.*, 2020)** (Figure. 7b). Co and Ni ions have similar chemical properties, so they work in synergy and play important role in plant metabolism.

These *TmONs* may have entered into root cells via apoplastic/synplastic process, ABC transporter, endocytosis, ion-carriers like IRT1, etc., where they get involved in various metabolic processes upon interaction with different organelles **(Hatami *et al.*, 2016)**. Various transporter proteins are present in plants for the translocation of respective metal ions. For instance, $Co^{2+}$ is transported from soil to the epidermis along with iron with the help of iron-regulated transportin proteins, IRT1, $Cu^{2+}$ by HMA5 and $Ni^{2+}$ by NRAMP (Figure.7b). Ferroportins like FPN1 and FPN2, are found in the plasma and vascular membrane, transfers these ions into vacuoles, causing sequestered in root cells. Before being transported to shoots, these ions forms complexes with citrate, histidine, methionine, or nicotinamide in the xylem. In leaves, these ions are released, and take part in metabolisms involving ABC transporter.

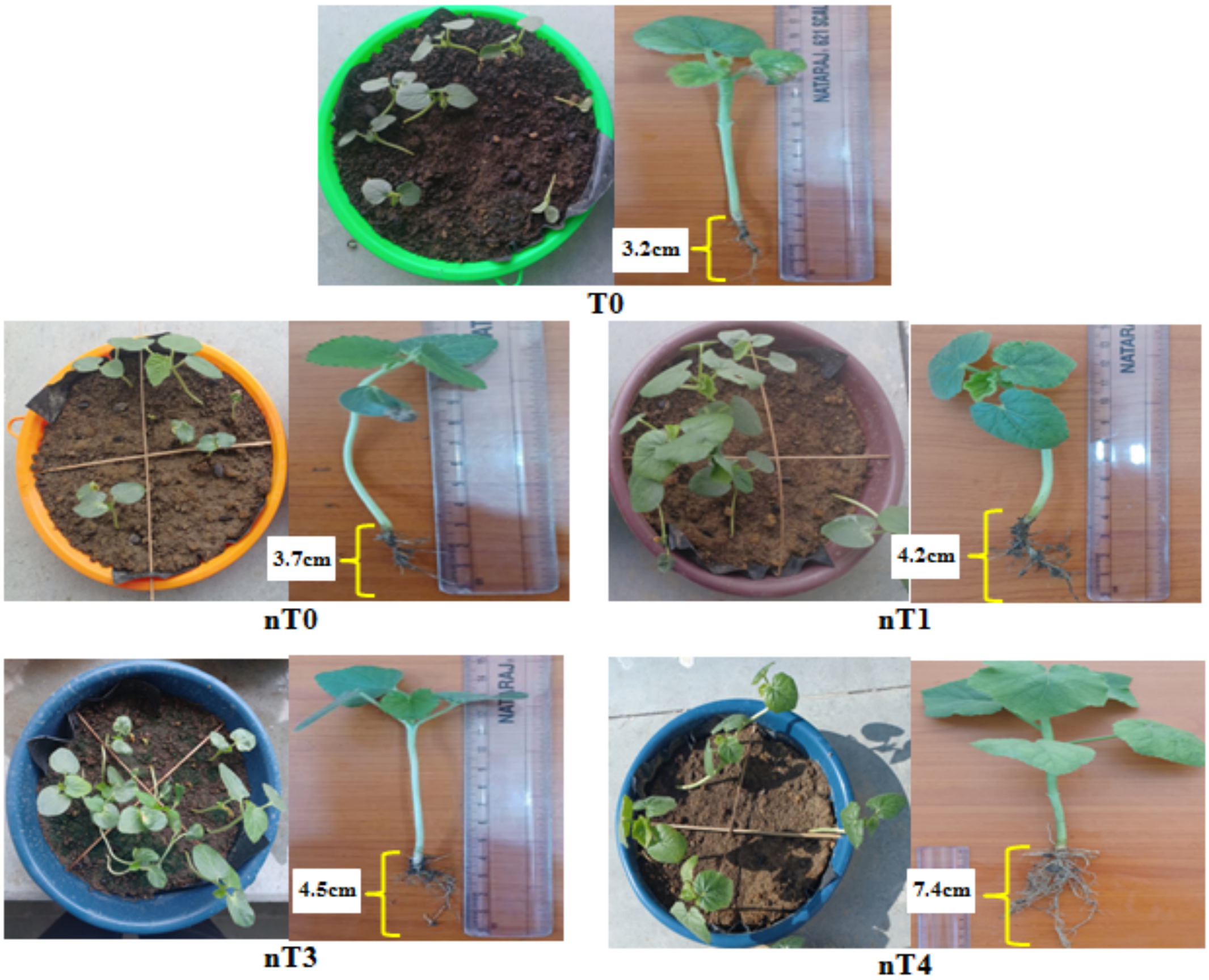

a)

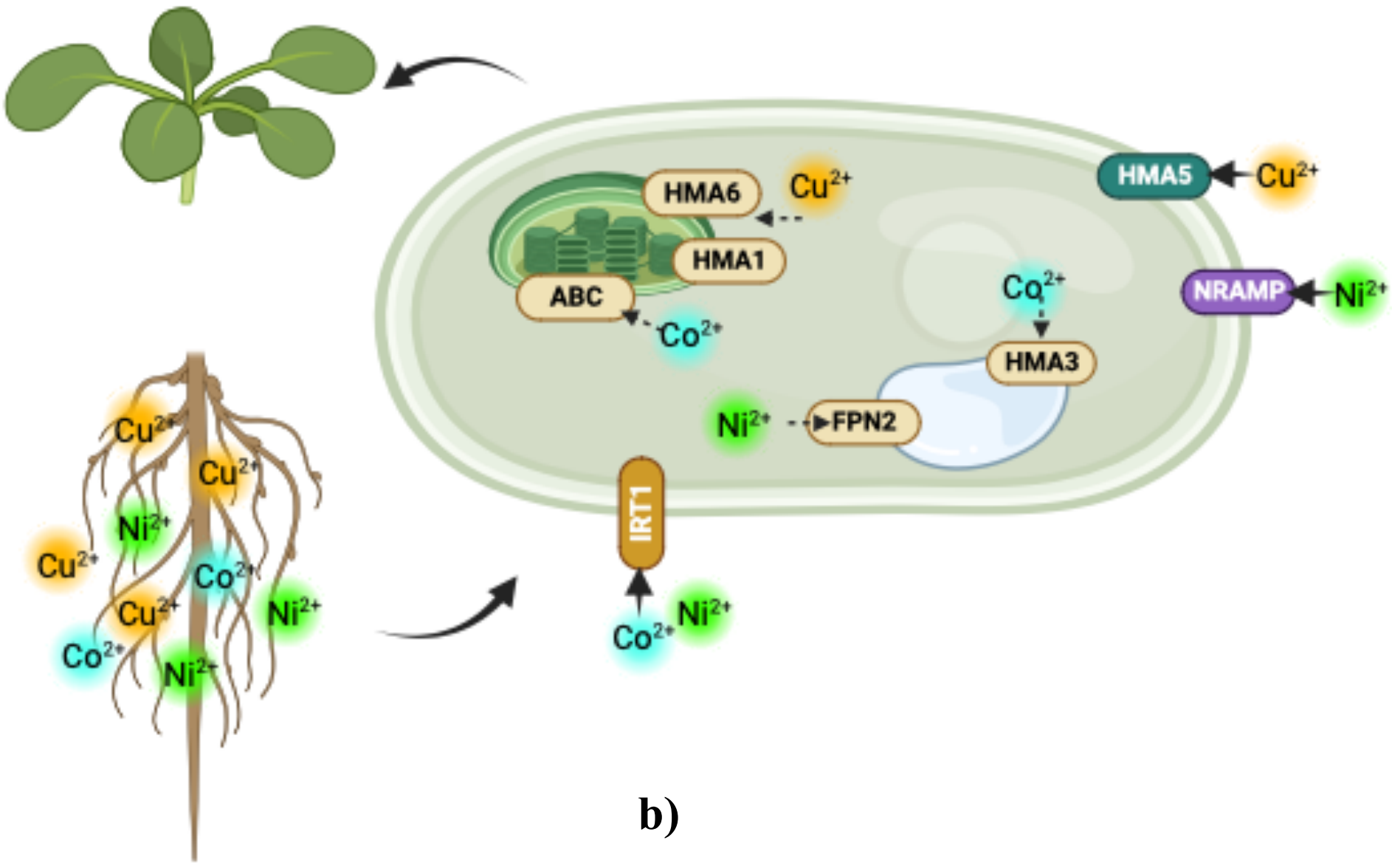

b)

**Figure. 7** Ionic pathway for the metal ions ($Cu^{2+}$, $Ni^{2+}$ and $Co^{2+}$) moving from roots to leaves. IRT1, NRAMP, HMAS, FPN2, HMA1/6/3, and ABC are transporter proteins involved in translocation.

As a consequence, deficiency of any of these metals mentioned in our work can lead to stunted growth, reduced root development, reduced seed production, changes in leaf shape, and decreased chlorophyll content. On the other hand, an excessive amount can have similar negative effects. Therefore, a balanced and controlled application of these nanoparticles is crucial for maximizing their benefits and minimizing potential adverse effects. It is worth mentioning that no such adverse effects were seen in the treated vermicompost plants during the whole study period, which may be related to the optimal *TmONs* concentration.

## 4. Materials and method

For the green synthesis of *TmONs,* the chemicals cupric acetate, Cu $(CH_3COO)_2$. $H_2O$, 98%, nickel acetate, Ni $(CH_3COO)_2$. $4H_2O$, 98%, cobalt nitrate, Co $(NO_3)_2$. $6H_2O$, 97% and ethanol, $C_2H_5OH$, 99.9% were used. Leaves of *M. oleifera* were obtained from the premises of Rajiv Gandhi Institute of Petroleum Technology, Jais, India. Okra seeds were purchased from local seed vendor of Jais, India. The earthworm species, *E. fetida* were used to bio-transform agricultural wastes into vermicompost at the Bioconversion and Engineering Laboratory, Rajiv Gandhi Institute of Petroleum Technology, Jais, India. A multi-elemental standard stock of 100 $\mu g/mL$, FINAR-92, provided by Inorganic ventures was used for preparation of standards of 1ppm, 5ppm, 50ppm and 100ppm. Ultrapure (Milli-Q®, Millipore, USA) water was used for preparing solutions and for dilutions during ICP-OES analysis.

### *4.1. Production of nano-organic vermicompost*

To prepare blended (RS+PM) vermicompost (T0-T3) as culture medium for plant growth, mature vermiproducts have been prepared with slight modification according to **Srivastava *et al.*, 2023**. Four distinct experimental treatments were mixed and made in the following ratios: cow dung (CD) as the control, rice straw (RS) (chopped small), and pressmud (PM):

Feedstock (T0) – CD (dry weight)

Feedstock (T1) – CD + RS in 1:1

Feedstock (T2) – CD + PM in 1:1

Feedstock (T3) – CD + RS + PM in 1:1:1

The *M. Oleifera* extract was prepared by decoction method as described by **Vongsak *et al.*, 2013**. 12 g of washed leaves were heated in 200 mL ethanol at 65°C and filtered. *TmONs* were fabricated using green solution combustion method, using the extract as fuel (60 mL) and salt precursors as oxidants in appropriate weight ratio ($3Cu^{2+}:2Ni^{2+}:1Co^{2+}$). The solution was stirred at room temperature for 8 hours. Afterwards it was kept in muffle furnace for combustion at 350°C. After washing the obtained powder was calcined at 600°C. In order to study the potency of nanoparticles, 1 kg of derived vermicompost was mixed with 100 ppm of biogenic *TmONs*. Furthermore, it was used with soil (1:4) to evaluate the growth parameters of plants. These mixtures were, negative control as T0 (without *TmONs*), positive control as nT0 (T0 + *TmONs*), nT1 (T1 + *TmONs*), nT2 (T2 + *TmONs*) and nT3 (T3 + *TmONs*).

### *4.2. Growth parameters of A. esculentus*

The germination study was performed on 30 seed sown in prepared soil mixes. They were allowed to germinate in open sunlight at ambient temperature. The seedlings were monitored at 10.00 am for 21 days in continuation. Completion of germination was indicated by radicle elongation to a minimum length of 2 mm. Moreover, the germination parameters namely, cumulative germination percentage (CGR), co-efficient velocity of germination (CVG), germination percentage (G%), mean germination time (MGT) and seedling vigor index (SVI) were calculated by using the formulae given **(Xu and Du, 2023 and Zhao *et al.*, 2021)**:

$$\text{CGR (\%)} = \sum_{i=1}^{n} \left(\frac{Ni}{Ts}\right) X\ 100$$

$$CVG\ (\%) = \frac{G1 + G2 + G3 + \cdots Gn}{(1 \times G1) + (2 \times G2) + \cdots n \times Gn} \times 100$$

$$G\ (\%) = \frac{\text{Total seed germinated at end of test}}{\text{Total number of initial seed}} \times 100$$

$$MGT\ (\%) = \frac{\sum(n \times d)}{N} \times 100$$

$$SVI = \text{Average of seedling length (cm)} \times G(\%)$$

where, *Ni* denotes the number of seed germinated on day *i*; *Ts*, total number of seed in each treatment; *i*, alternate day after start; G, germination on each day after start; *n*, number of seeds germinated on each day, d, number of days from the start of the test; N, total number of seeds germinated at the experiment's completion.

*4.3. Chemical analysis and analytical measurements*

*Vermicompost analysis*: The physicochemical parameters namely, pH and electrical conductivity (EC) were recorded by a digital pH and electrical conductivity meter. The loss-on ignition method was used to analyse total organic carbon (TOC) in a muffle furnace. Acid digestion and Kjeldahl distillation were used in succession to calculate total nitrogen (TN) content. Total phosphorus (TP) and total potassium (TK) were each quantified using a spectrophotometer (di-acid digestion) and a flame photometer, respectively as described in our previous work (**Srivastava et al., 2023; data available in supplementary material**).

*TmONs characterization*: The physicochemical characterizations of the synthesized nanoparticles were carried out using FTIR (Thermoscientific, NICOLET Is20) between the wavelength range of 4000 and 400 $cm^{-1}$ to know about the functional groups present, UPLC-PDA (Waters Acquity UPLC, H-class system) was done to determine the poly phenols and phenolic acids present in the extract. The crystal structure was analysed using XRD (PANalytical) equipped with parallel beam geometry with Cu Kα radiation (λ=1.5406 Å) with a voltage of 45 kV, current 40 mV and angles ranged from 10° to 90° in 0.003 steps. Oxidation states of the metal oxides nanoparticles were determined using XPS (Thermoscientific, K-ALPHA$^+$ Surface Analysis) with an x-ray source Al Kα (Monochromatic) with 6 mA beam current and 12 kV, FESEM (JEOL, JSM-7900F) at an accelerating voltage of 5-10 kV, and HR-TEM (JOEL JEM 2100) at a voltage of 200 kV were used to obtained the images of the synthesized nanoparticles.

*Heavy metal quantification*: For toxicological risk assessment of heavy metals, ICP-OES (Thermo Fisher Scientific, iCAP pro X Duo) at plasma flow 15 L/min, pump rate 15 rpm and rinse time 30 s with QTEGRA software was used for metal quantification. Wavelengths for Cu, Ni and Co were chosen as 324.754, 231.604, 238.892 nm respectively. For soil, triple acid digestion process was performed according to **Senila *et al.* (2011)**. In brief, 0.1 g of sample was mixed with a reagent HF: $HNO_3$:HCl:$H_2O$ (3:3:1:3) and the solution was heated till dryness. Thereafter, 3 mL of aqua-regia was further added till a clear solution was obtained. The remain was made up to 25 mL using Milli-Q water with 1% $HNO_3$. For Okra leaves, open vessel hot method was used as described by **Senila *et al.* (2011)**. 0.5 g of leaves powder was added to 10 mL $HNO_3$, and the mixture was heated for 3 hr, then was cooled and further $H_2O_2$ was added till clear solution was obtained. After digestion, the samples were filtered and made-up to 30 mL for leaves sample. Pre-cleaned PTFE tubes were used to put digested samples until the quantification was done using ICP-OES.

*Toxicity test on E. fetida*: Prior to application of synthesized nanoparticles, the toxicity assessment was also conducted ascertaining the survival of earthworm. A concentration of 50, 100, 200, 300 and 500 ppm of derived *TmONs* were used with 1 kg of cow dung to perform toxicity assay. A three-week study was performed to assess the impacts of *TmONs* on *E. fetida* by monitoring the survival to visually evaluate the toxicity and risk.

*4.4. Statistical analysis*

The variations in physicochemical parameters between various feedstocks were statistically examined ($P < 0.05$) using one-way analysis of variance. Duncan's multiple range tests and Tukey's b as post hoc tests were used to examine the variations in chemical attributes between the original feed combinations and the final vermicompost.

## Conclusions

Novel biogenic trimetallic (Cu/Ni/Co) oxide nanoparticles derived from *Moringa oleifera* leaves extract was successfully synthesised using green solution combustion method. When combined with vermicompost, these nano-oxides have the capacity to bind or decompose a variety of soil pollutants, including organic contaminants and heavy metals that help with soil remediation. Overall positive effects on the growth parameters of Okra plants resulted in a significant increase in cumulative germination, co-efficient velocity of germination and SVI by 167%, 67% and 95%, respectively while reduction in MGT by 41% in comparison to the control group. The electrical conductivity, total organic carbon and NPK content in produced vermicompost were found to be significantly ($P < 0.05$) increased by 31%, 16%, 26%, 20% and 72% respectively due to the effect of PM and RS. The carbon to nitrogen (C/N) ratio dropped by 11.1%. Moreover, ecological risk assessment of TmONs concentration has also been evaluated and found to be under maximum permissible limits of WHO (1996). This synergetic approach for nano-organic amendment continues to serve a dual purpose, recycling agro-wastes and at the same time impact the agricultural transformation by using eco-friendly fertilizer. The findings of this research will contribute to the knowledge base and assist farmers and researchers in making further related studies regarding the use of these amendments especially the neglected metal ions like Co to promote soil health, nutrient availability, and plant productivity. However, this approach is new and needs further in-depth studies in regard to the variability and reactivity of nanoparticles in different environments.

**Supplementary data**

E-supplementary data for this work can be found in e-version of this paper online.

**CRediT authorship contribution statement**

**Shalu Yadav**: Conceptualization, Methodology, Writing, Reviewing and Editing, Data verification, **Ankit Singh:** Data Curation, Reviewing and editing, **Praveen Kumar Srivastava**: Conceptualization, Methodology, Formal Analysis Software, Data Curation, Writing- Original draft preparation, **Abhay Kumar Choubey**: Reviewing and Editing, Data Verification, Validation, Supervision.

**Declaration of competing interest**

The authors declare that they have no known competing financial interests or personal relationships that could have appeared to influence the work reported in this paper.

**Acknowledgments**

The authors are thankful to Rajiv Gandhi Institute of Petroleum Technology, Jais, Amethi - 229304, India for providing the necessary facilities during the research.

**Funding**

This research did not receive any specific grant from funding agencies in the public, commercial, or not-for-profit sectors.